\documentclass{emulateapj}

\shorttitle {Ultra deep sub-kpc view of nearby massive compact galaxies} 
\shortauthors {I. Trujillo, E. R. Carrasco and A. Ferr\'e-Mateu}

\begin{document}

\title {Ultra deep sub-kpc view of nearby massive compact galaxies}

\author{Ignacio Trujillo\altaffilmark{1,2}}  
\affil{Instituto de
Astrof\'isica de Canarias, V\'ia L\'actea s/n, 38200 La Laguna,
Tenerife, Spain}
\email{trujillo@iac.es}

\author{Eleazar R.  Carrasco}  \affil{Gemini Observatory/AURA, Southern
Operations Center, Casilla 603, La Serena, Chile} 

\and

\author{Anna Ferr\'e-Mateu\altaffilmark{2}}  \affil{Instituto de
Astrof\'isica de Canarias, V\'ia L\'actea s/n, 38200 La Laguna,
Tenerife, Spain}

\altaffiltext{1}{Ram\'on y Cajal Fellow}
\altaffiltext{2}{Departamento de Astrof\'isica,
Universidad de La Laguna, E-38205 La Laguna, Tenerife, Spain}

\begin{abstract}

Using  Gemini North telescope ultra deep and high resolution (sub-kpc) K-band adaptive optics imaging of
a sample of 4 nearby (z$\sim$0.15)  massive ($\sim$10$^{11}$$M_{\sun}$) compact (R$<$1.5 kpc) galaxies,
we have explored the structural properties of these rare objects with an unprecedented detail. Our
surface brightness profiles expand over 12 magnitudes in range allowing us to explore the presence of any
faint extended envelope on these objects down to stellar mass densities $\sim$10$^{6}$
M$_{\sun}$/kpc$^{2}$ at radial distances of $\sim$15 kpc. We find no evidence for any extended faint tail
altering the compactness of these galaxies. Our objects are elongated, resembling visually S0 galaxies,
and have a central stellar mass density well above the stellar mass densities of objects with similar
stellar mass but normal size in the present universe. If these massive compact objects will eventually
transform into normal size galaxies, the processes driving this size growth will have to migrate around
2-3$\times$10$^{10}$$M_{\sun}$ stellar mass from their inner (R$<$1.7 kpc) region towards their
outskirts. Nearby massive compact galaxies share with high-z compact massive galaxies not only their
stellar mass, size and velocity dispersion but also the shape of their profiles and the mean age of their stellar
populations. This makes these singular galaxies unique laboratories to explore the early stages of the
formation of massive galaxies.

\end{abstract}


\section{Introduction} 
\label{Introduction}

Following the discovery (Daddi et al. 2005; Trujillo et al. 2006) that massive spheroid-like
galaxies (M$_\star$$\gtrsim$10$^{11}$$M_{\sun}$) at z$\gtrsim$1 were significantly more
compact that their local equivalent counterparts, there have been some efforts exploring
whether any single of these galaxies can be found in the nearby Universe (Trujillo et al.
2009; Taylor et al. 2010; Valentinuzzi et al. 2010; Shankar et al. 2010, Shih \& Stockton 2011). According to some
theoretical models (Hopkins et al. 2009a) it would be possible that some of the massive
compact high redshift galaxies should have survived untouched since their formation epoch.
Consequently, the discovery of a population of nearby old compact massive galaxies would 
open the possibility of exploring the galaxy formation mechanisms of the early universe in
exquisite detail. So far, the amount of nearby massive galaxies with sizes ($\lesssim$1.5
kpc) similar to the median value   found at z$\gtrsim$2 (e.g. Trujillo et al. 2007; Buitrago et al.
2008) represents only a tiny fraction ($\sim$0.03\%) of the massive objects in the nearby
universe (z$<$0.2; Trujillo et al. 2009). Moreover, contrary to theoretical expectations,
the nearby compact massive galaxies have average stellar ages which are relative young
($\sim$2 Gyr), more younger than the age expected if they were relics from an early formation
epoch. In this sense, present day massive compact galaxies, more than remnants from an early
epoch, resemble  almost perfect counterparts of the massive galaxy population found at
z$\sim$2: they have similar stellar masses, sizes and ages (Trujillo et al. 2009; Ferr\'e-Mateu et al.
2012). What is currently missed though, is
an in-depth analysis of the morphological properties of the nearby massive compact
population.

Detailed analysis of the stellar mass density profiles of massive compact galaxies at high redshift
(Bezanson et al. 2009; Hopkins et al. 2010; van Dokkum et al. 2010; Carrasco et al. 2010) show that these
objects have a moderate excess of stellar mass density at their centers and a significant lack of stars
in their outer regions. In fact, it is this lack of a tail in their stellar profiles what is making them
to look so compact. The studies of the morphological properties of the nearby massive galaxy population
have been seriously limited as the presently available imaging of these objects is based on ground based
observations (with a typical seeing of $\sim$1 arcsec or equivalent to $\sim$2.6 kpc at z=0.15). This
has prevented a detailed analysis of the inner region of these galaxies. Also, the outer parts of these
galaxies have not been studied in detail. For that reason, a deep sub-kpc view of these galaxies would be
of great help to put these galaxies into context within the local massive galaxy population. Under the
assumption that these massive compact galaxies are similar objects to those found at high redshift (i.e.
both in structural and stellar population properties) the unprecedented detailed analysis that can only
be conducted in nearby galaxies will allow us to constrain the different evolutionary paths that transform these
galaxies into the massive "normal sized" galaxy population. In particular, with this work we would like
to answer the following questions: what is the morphological
nature of the massive galaxies in their primitive state: disky or spheroidal? Does the
compactness nature of these objects an artifact due to the missing of light in the outer
regions?  In this paper, we present ultra-deep high resolution K-band imaging obtained with
the Gemini North telescope of a sample of four nearby massive compact galaxies from the sample
presented in Trujillo et al. (2009). We will show that the most common nature of the nearby
massive compact galaxies is disky and that there is not any evidence indicating that their
stellar mass density profiles have an extended outer envelope biasing their size estimates.
In what follows, we adopt a cosmology of $\Omega_m$=0.3, $\Omega_\Lambda$=0.7 and H$_0$=70 km
s$^{-1}$ Mpc$^{-1}$.

\section{The Data} 
\label{TheData}

The galaxies studied here are a sub-sample of the collection of nearby compact massive
galaxies  compiled by Trujillo et al. (2009). The original sample contains 48 bona-fide
compact massive  galaxies taken from the New York University Value-Added Galaxy Catalog
from the SDSS Data  Release 6 (Blanton et al 2005; NYU from now onwards).  Our galaxies  have a mean redshift of
0.16, a mean effective  radius of $\sim 1.3$ kpc  with no detected signatures of AGN in
their spectra which might  affect the size determination, and a mean stellar mass of $\sim
9.2 \times  10^{10}$  M$_{\odot}$ (Chabrier 2003 Initial Mass Function). 

The high spatial resolution imaging presented here was obtained with  the Gemini North
telescope using the Near-Infrared Imager and Spectrometer  (NIRI; Hodapp et al. 2003) with
the ALTAIR/LGS (Laser Guide Star) adaptive optics  systems  (Herriot et al. 2000, Boccas et
al. 2006). ALTAIR requires a relatively bright  star ($\lesssim$ 18 mag in R-band) in the
proximity of the target object (within 17\arcsec)  to  obtain a good Strehl correction.  We
got time to explore four galaxies of the Trujillo's sample that satisfy this criteria. The
final galaxy sample has Adaptive Optic (AO) tip-tilt stars with magnitudes  between 15.8 and
17.9 in R-band located within 16\arcsec of the main target. 

The galaxies were observed during the first semester of 2010, in queue mode, using the  K
(2.2 $\mu$m) filter and NIRI f/14 camera, which provides a field of view of  51\farcs1
$\times$ 51\farcs1 with pixel scale of 0\farcs0488 on side onto a  1024 $\times$ 1024 ALADDIN
InSb array.  In addition, standard stars  for photometric calibration were observed before or
after our galaxies. The  standard stars were used to determine the photometric zero points
and monitor the  image quality of the observations. Throughout the standard stars and the
field stars presented in our observations we determined the effective FWHM of our
observations. From standard stars, the effective FWHM varied between 0\farcs11 and
0\farcs13  (note that the photometric standard stars have always the maximum Strehl
corrections because they are used as AO tip/tilt star). From field stars,  the effective
FWHM  was between 0\farcs16 and 0\farcs24 varying with the stellar magnitude, and the 
location of the star relative to the galaxy of interest. The galaxy 
SDSS-J153934.07$+$441752.2 was observed in separate nights (see Table~1), but at similar
airmass. For galaxy J120032.46+032554.1 only 27 images of 60 sec were observed (half of the
planned observations). Despite this reduction, the combined image for this galaxy is still
 very deep allowing us to reach faint surface brightness.

The data were processed following the standard procedures for near-infrared
imaging using the NIRI/Gemini IRAF package v1.10. Normalized flat field images were 
constructed from flat images observed with the Gemini Calibration unit (GCAL) with the
shutter closed (lamps off) and shutter open (lamps on).  Dark images observed at the end
of each night and flat field images (lamp off) were used to construct a bad pixel mask with 
bad and hot pixels.  The sky images were constructed from the raw science images by 
identifying objects in each frame, masking them out, and averaging the remaining good
pixel (the images were observed with a 3\arcsec\, $\times $ 3\arcsec\, mosaic pattern).
The raw science images were then processed by subtracting the sky on a frame-by-frame basis
and divided by the normalized flat field images. Finally, the processed images were registered
to a common pixel position and median combined. The final images have a field of view of 
39\farcs5 $\times$ 39\farcs5. The images of our four galaxies are shown in Figure~1.

Photometric calibrations were derived using the UKIRT Mauna Kea Observatories JHKL'M
Standard Stars FS 132 (s860-d), FS 152 (p460-e) and P272-D (Leggett et al. 2006)
and the P064-D faint standard star (Persson et al 1998). Given that only one standard star 
was observed for each galaxy (before or after), we have used an average value for the 
extinction of $k_{K} = 0.052 \pm 0.028$ from Leggett et al. (2006). Table~1
lists all observational parameters for our NIRI observations. Since we use an average
value for the extinction coefficient, the estimated  error in the photometric calibration 
will be driven by the uncertainty in the correct value of the extinction coefficient for
the night of observation. Hence, we estimated the error by summing in quadrature the 
median error of the aperture photometry, the error of the standard catalog and
the median error of the extinction coefficient. The error varied between 0.03 mag
and 0.05 mag, depending on the star. For each galaxy, the adopted zero point is listed
in the column (7) of Table~1. The values were transformed from the Vega system to 
the AB system using the relation $K_{AB} - K_{Vega} = 1.91$ mag.

\section{Analysis} 
\label{Analysis}

A visual inspection of the nearby compact massive galaxies shown in Fig. 1 indicates that the most common
morphology of our objects is disky. In fact, two objects (SDSS J103050.53$+$625859.8 and SDSS
J120032.46$+$032554.1) visually resemble S0 galaxies viewed in edge-on projection. The other two galaxies
(SDSS J153934.07$+$441752.2 and SDSS J212052.74$+$110713.1) have a more distorted morphology but still
are compatible with being S0 galaxies with a lower inclination (see also Valentinuzzi et al. 2010). On what follows, we make a quantitative
analysis of  the structural properties of the nearby compact massive galaxies.

\subsection{K-band surface brightness profiles}

In Fig. 2 we show  circular aperture K-band surface brightness profiles of our compact galaxies. Although
our galaxies have a clear elongation, circular apertures are used to allow a direct comparison with the
circular averaged profiles of "normal-sized" galaxies as we will show later. To create our profiles we
followed the same technique explained in Pohlen \& Trujillo (2006). Briefly, we obtain first the surface
brightness profile of the galaxy up to large distances. Then we estimate the sky contribution in those
regions outside the galaxy where the profile is flat and we remove (add) this value from (to) the images.
After doing that, we calculate again the surface brightness profiles of these galaxies.

For most of our galaxies we can explore their surface brightness profiles down to $\sim$27 mag/arcsec$^2$. This implies
probing around 12 magnitudes from the peak of their surface brightness distribution down to their last observed points. This
extraordinary depth allow us to investigate whether there is any evidence for any extra hide (halo-like) component in our
galaxies which due to its faintness could not have been observed in previous shallower images of these objects. We can
consequently address whether earlier works have  incorrectly measured the sizes of these objects towards lower values.

We have fitted our profiles with a single S\'ersic (1968) model to compare the sizes we get from our images in relation to those
obtained in previous shallower and worst resolution SDSS data (Trujillo et al. 2009). We avoided in this fitting those
regions of the galaxies more severely affected by the PSF (Point Spread Function). This means we only take those points
beyond the FHWM of our PSF (this distance is indicated by a vertical dashed line in Fig. 2). The effective radii and
S\'ersic indeces we got are shown in Table 2. We explored whether our results were affected by this particular radial range
of exploration. To do that we repeat our fitting only taking the points outside 2 times the FWHM. Our estimates remained
very well constrained with changes in the effective radius below 15\% and for the S\'ersic index less than 11\%. This
robustness is due to the extreme depth of our images.

We can now compare these values against the sizes obtained using SDSS imaging. In Trujillo et al. (2009)
we got the following results: 1.42 kpc (103050.5), 1.31 kpc (120032.4), 1.11 kpc (153934.0) and 1.38 kpc
(212052.7). We can see that the agreement with the present much deeper and higher resolution data is
excellent with differences less than 7\%. This is an indirect proof that there is not any hidden component in these massive nearby compact
galaxies that could alter the size of the objects. The main novelty that the present deep data allow us
to explore is the shape of these galaxies. We observe that these objects are well fitted with moderately
low S\'ersic indeces (2$<$n$<$4) values. The absence of large S\'ersic indeces is again against the idea
that there is a missing faint component surrounding these objects. 

\subsection{Stellar mass density profiles}

In order to understand the building of the massive galaxies it is worth comparing the stellar mass
density profiles of the compact population against the mass distribution of galaxies of a similar mass
but with normal sizes. We have transformed our observed K-band surface brightness profiles into stellar mass
density profiles using the total stellar masses measured in  Blanton et al. (2005)  listed in
Table 2. We have assumed that the stellar mass to light ratio is constant along the radial distance of the
galaxy. The outcome of this exercise is presented in Fig. 3.

To build the stellar mass density profiles of the "normal-sized" galaxies used as a reference, we took the
structural parameters (S\'ersic index n, effective radius r$_e$ and stellar mass M$_\star$) of all the galaxies in the NYU
catalogue (Blanton et al. 2005) with 0.8$<$M$_\star$$<$1.2$\times$10$^{11}$M$_{\sun}$ and 0.1$<$z$<$0.2. These NYU structural
parameters were retrieved from profiles obtained using circular apertures. To facilitate the comparison
with our  profiles, we divided the  NYU galaxies into two different categories: disk-like (n$<$2.5)
and spheroid-like (n$>$2.5). We find that the average disk-like massive galaxy within the NYU sample at
those redshifts has M$_\star$=0.95$\times$10$^{11}$M$_{\sun}$, n=2 and r$_e$=5.7 kpc. On the other
hand,  the average spheroid-like object has M$_\star$=0.98$\times$ 10$^{11}$M$_{\sun}$, n=4 and r$_e$=4.7
kpc.

Once we obtained these average galaxy profiles, the representative regions of each galaxy
category were build using all the galaxies in the NYU sample within the above stellar mass range and redshift
interval whose central stellar mass densities were within the 68\% of the distribution centered around the mean
value.

From the comparison of the stellar mass density profiles of the compact galaxies with those of
"normal-sized" objects (Fig. 3) is straightforward to conclude that the compact galaxies do not resemble
neither of the two categories. Although visually the elongation of the compact galaxies would suggest
that these objects are more likely disks, the shape of these profiles are closer to those considered as
spheroids in the local universe. It is easy to see that there is  an excess of mass at the center of the
compact galaxies and a lack of stars (starting mainly around 3 kpc) in the outer regions. The deep
profiles that we present here undoubtedly show that nearby massive compact galaxies do not have an extended
outer component and, consequently, are genuinely compact.

An interesting exercise that can be conducted is to estimate the amount of stellar mass within the inner
(R$<$3 kpc) region of the compact galaxies and compare this to the "normal-sized" objects. In the case of
the compact galaxies we find that the stellar mass fraction inside 3 kpc ranges from 0.72 (in the case of
212052.7) up to 0.89 (for 153934.0). In the case of "normal-sized" objects these fractions are
significantly lower: 0.27 for disk-like objects and 0.38 in the case of spheroids. This implies that
there is $\sim$2 times more stellar mass inside 3 kpc in the case of the compact massive galaxies than in
objects of the same stellar mass but normal size. This difference in stellar mass implies that one would
expect a much larger  central velocity dispersion in the case of compact galaxies compared to  normal
galaxies with equivalent stellar mass. A crude estimation (following the virial theorem expectation)
suggests that this increase should be of the order of $\sqrt2$ as there is a factor of 2 more stellar
mass in the central regions. Are these expectations in agreement with observations? In Trujillo et al.
(2009) we found that the average central velocity dispersion of 10$^{11}$M$_{\sun}$ galaxies according to
SDSS was 180 km/s. Our massive compact galaxies have (see Table 2) an average $\sigma$=243 km/s. This is
1.35 larger than the value found in normal galaxies of the same stellar mass and fits very well with the
virial $\sqrt2$ expectation. This is another indirect proof of the larger stellar mass densities that
massive compact galaxies have in their centers.

\section{Discussion}

\subsection{Can massive compact galaxies be transformed into the core of giant ellipticals?}

A popular idea is that  massive compact galaxies at high-z will end being the central part of present day most massive
objects (Bezanson et al. 2009; Hopkins et al. 2009b). This scenario is supported by many indirect observational evidences as
the progressive growth of the wings of the profiles of the massive galaxies with time (van Dokkum et al. 2010), the larger
velocity dispersion of the massive galaxies at high-z ($\sim$1.5 times larger) compared to equally massive objects today
(e.g. Cenarro \& Trujillo 2009, Cappellari et al. 2009, Onodera et al. 2010, Newman et al. 2010, van de Sande et al. 2011),
the similar number density ($\sim$2$\times$10$^{-4}$ Mpc$^{-3}$) between 10$^{11}$M$_{\sun}$ massive compact galaxies at
high-z and today $\sim$2 times more massive objects (van Dokkum et al. 2010, Cassata et al. 2011, Buitrago et al. 2012), the
expected mass growth by a factor of $\sim$2 of the massive galaxies with time expected theoretically (Naab et al. 2009,
Sommer-Larsen \& Toft 2010, Feldmann et al. 2010, Oser et al. 2012) and suggested by the observations (see e.g. Trujillo et
al. 2011). In this sense, it is natural to compare our stellar mass density profiles with the profiles of normal-sized
objects in the local universe but with a stellar mass twice the ones found for the local massive compact galaxies. The
unprecedented resolution of our profiles  allow us to see whether there is any change at sub-kpc level in the structure of
these objects. In particular, we are interested on estimating how the growing processes that could eventually bring the
compact galaxies into the core of more massive objects affect  their inner regions. Moreover, we would like to quantify
which number of stars should migrate during such transformation towards the outskirts of these objects.

In Fig. 4 (left panel) we show the comparison between the stellar mass density profiles of our nearby compact massive
galaxies against objects with normal sizes  but 2 times more massive. The nearby massive compact galaxies
are clearly more dense at the center than the objects they will potentially be transformed in the future.
That implies that the transformation from one class to another should imply a significant migration of
the stars from the center of the compact galaxies towards their outer regions. The stellar mass density of both
normal-sized and compact galaxies are similar at R=1.7 kpc. We can consequently estimate which is the
excess of stellar mass of the compact galaxies compared to the other objects within this radius. We find
that compact galaxies have  $\sim$3$\times$10$^{10}$M$_{\sun}$ more mass than a disk-like normal size
galaxy and $\sim$2$\times$10$^{10}$M$_{\sun}$ more mass than a spheroid object within 1.7 kpc. That means such
amount of stellar mass should be relocated outwards of 1.7 kpc after the evolution of the compact galaxy
into a larger size object. This enormous evolution of the inner region of the compact galaxies is
expected in a minor merging scenario (see e.g. Fig. 3 of Oser et al. 2012). Both compact massive galaxies
and the larger objects have the same amount of stellar mass ($\sim$8$\times$10$^{10}$M$_{\sun}$) within
$\sim$5 kpc. This suggests that the processes responsible of the growth of the galaxies locate most of
the new assembled stars ($\sim$10$^{11}$M$_{\sun}$) beyond that radius.

At present it is still unclear which exact mechanism is ultimately responsible of the growth of the massive compact
galaxies. The most favoured scenario is the minor dry merging scenario (e.g. Khochfar \& Burkert 2006; Maller et al. 2006;
Hopkins et al. 2009a; Naab et al. 2009; but see Nipoti et al. 2012 for a critical vision) but there has been also some support to the puffing-up model. In this last picture
(Fan et al. 2008; 2010), galaxy growth is connected to the removal of gas as a result of AGN activity or stellar activity
(Damjanov et al. 2009). Based on recent simulations (Ragone-Figueroa \& Granato 2011) the growth associated to this process
should be very fast ($\sim$20-30 Myr). This would imply that objects with stellar populations older than $\sim$1 Gyr should
already have normal sizes. In which scheme better fit the properties of our nearby massive compact galaxies?

Our nearby massive compact galaxies are young but they stellar populations are older than 1 Gyr. This seems hard to
reconcile with the puffing up model. On the other hand, some of our objects seem to be accreting minor satellites as they
have distorted morphologies. Not having spectroscopic redshifts for the rest of the galaxies in our fields, we are unable to
quantify the presence of satellites around our galaxies (see e.g. Newman et al. 2012; M\'armol-Queralt\'o et al. 2012). So
far, the data obtained for the nearby compact galaxies seem to support more the minor merging scenario than the puffing up
model.

\subsection{Distant and nearby massive compact galaxies}

An interesting aspect that we can address with our data is whether our stellar mass density profiles resemble those of
equally massive compact galaxies at z$\sim$2. Contrary to what we find in the local universe, compact massive galaxies were
very common at those redshifts. Consequently, if we prove that our nearby massive compact objects are similar in structure
to their high-z counterparts, this would favor the idea that these objects share a similar formation origin. So far, not
only the size, stellar mass and velocity dispersion of the nearby compact galaxies are the same than their equivalent high-z
objects but  there is some evidence  implying that the compact nearby galaxies are  young too (with a mean stellar age of
$\sim$1-2 Gyr; Trujillo et al. 2009, Ferr\'e-Mateu et al. 2012). This age is equivalent to the age of the compact objects
found at high-z (e.g. Kriek et al. 2009). The question that we address in this paper is whether the distribution of the
stars in the nearby compact massive galaxies also follow a similar shape to the ones found at high-z. This is shown in Fig.
4 (right panel).

The objects that we have used to compare with our compact galaxies are the massive compact galaxies from
the sample of Szomoru et al. (2012) at z$\sim$2. This is a collection of 12 galaxies observed with the
HST WFC3 as part of the CANDELS (Grogin et al. 2011; Koekemoer et al. 2011) GOODS-South. The high-z objects have a median stellar mass of
8.3$\times$10$^{10}$M$_{\sun}$ (Kroupa IMF) and a median size of 0.84 kpc. The high-z profiles presented
in Szomoru et al. (2012) are deconvolved so we can have a better idea of how the profiles at high-z look
in their inner (R$<$1 kpc) regions. We can see that the agreement between the nearby and high-z compact
galaxies is excellent.

It is evident from the images shown in Fig. 1 that our galaxies are elongated. We have measured their
axis ratio using the ELLIPSE from IRAF. The axis ratio measured in the outer regions is shown in Table 2.
Is there any evidence about a similar elongated shape for the compact massive galaxies at high-z? van der
Wel et al. (2011) and Buitrago et al. (2012) have addressed this issue and find that this is the case.
Most of high-z massive galaxies at high-z have elongated shapes too. Summarizing, nearby massive compact
galaxies are almost exact copies of the high-z massive compact galaxies suggesting that their detailed
study can shed light on the formation of the first massive galaxies assembled in the Universe.

\section{Conclusions} 
\label{Conclusion}

Using ultra-deep imaging at a sub-kpc resolution of a sample of 4 nearby massive compact galaxies, we have
shown unequivocally that these objects are genuinely compact with no evidence of an extended faint
component altering their size estimate. These nearby massive compact galaxies have an elongated shape
resembling the structure of S0 objects. Their stellar mass density profiles are significantly more dense
in their inner regions than any  galaxy with similar stellar mass and normal size in the local universe. 

Nearby massive compact galaxies share a large number of properties with massive compact galaxies in the
distant universe: same stellar mass, size, shape, velocity dispersion and  the mean age of their
stellar populations. Contrary to high-z compact massive galaxies which were very common among the family
of massive galaxies at z$\sim$2, nearby massive compact galaxies are a tiny fraction of the family of
massive galaxies today. It still unclear whether the nearby compact massive galaxies are relics of that
epoch (which have had a recent burst of star formation that has rejuvenated their ages) or whether they
are truly young galaxies recently assembled. Undoubtedly, the study of these rare nearby objects open the
possibility of exploring in exquisite detail the early stages of massive galaxy formation.

\acknowledgments 

We thank the referee for his/her positive report. I.T. is  Ram\'on y Cajal Fellows of the Spanish Ministry of Science and
Innovation. This work has been supported by the Programa Nacional de Astronom\'ia y Astrof\'isica of the Spanish Ministry of
Science and Innovation under grant AYA2010-21322-C03-02. Based on observations obtained at the Gemini Observatory, which is
operated by the  Association of Universities for Research in Astronomy, Inc., under a cooperative agreement  with the NSF on
behalf of the Gemini partnership: the National Science Foundation (United  States), the Science and Technology Facilities
Council (United Kingdom), the  National Research Council (Canada), CONICYT (Chile), the Australian Research Council
(Australia),  Minist\'erio da Ci\^encia, Tecnologia e Inova\c c\~ao (Brazil) and Ministerio de Ciencia, Tecnolog\'ia e 
Innovaci\'on Productiva (Argentina). Program ID: GN-2010A-Q-24.

\facility{Gemini-South}{NIRI/Altair}

{}

\begin{figure}
%\epsscale{2.0}
\includegraphics[angle=0,width=\textwidth]{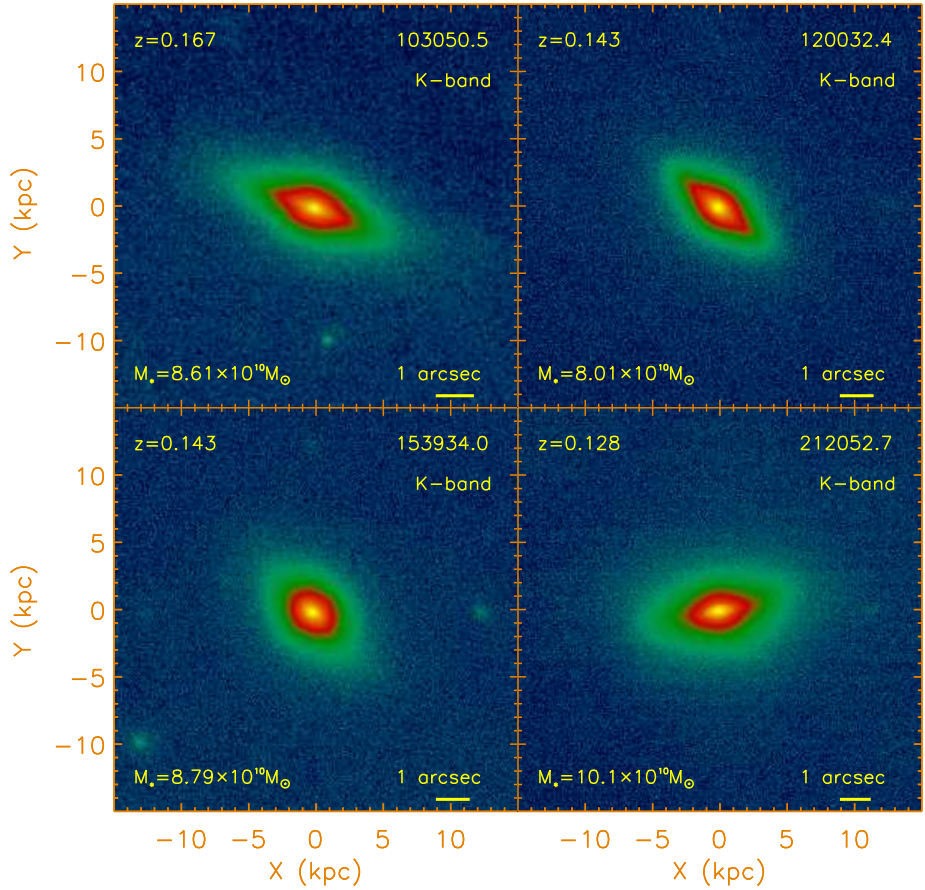}
%\plotone{figuracompactas.eps}

\caption{K-band Gemini high resolution (FWHM$\sim$0\farcs2) imaging of four nearby
(z$\sim$0.15) massive compact galaxies. Listed on each figure is the galaxy name, its stellar
mass and its spectroscopic redshift. The solid line indicates 1 arcsec angular size.}

\label{fig1}
\end{figure}

\begin{figure}
%\epsscale{2.0}
\includegraphics[width=\textwidth]{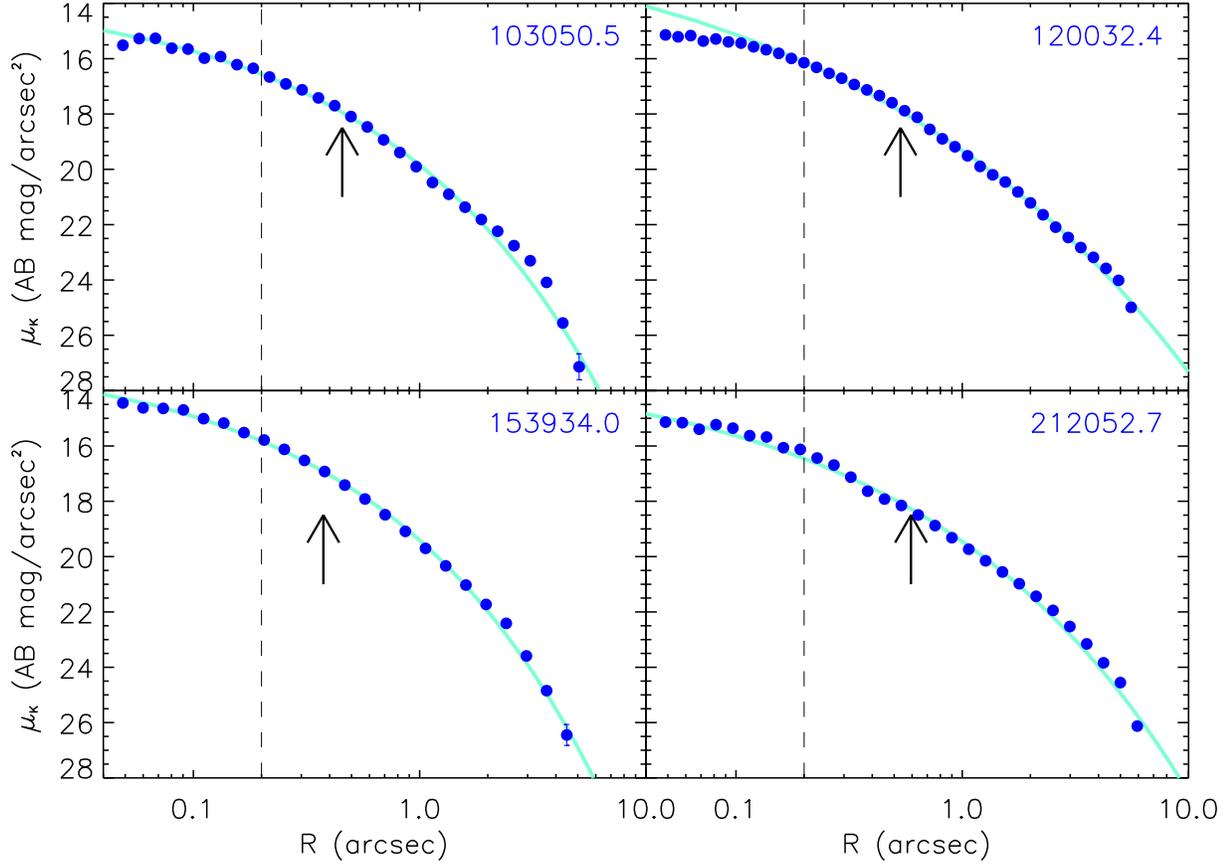}

%\plotone{allmassprofile.ps}

\caption{K-band surface brightness  profiles of our sample of nearby massive compact galaxies (blue
points). The soft blue lines are the best S\'ersic fit to the data. The vertical line shows a FWHM PSF of 0.2 arcsec (the typical resolution of our images). The
depth and high resolution of our images allow us to explore the profiles of our sample around 12
magnitudes in range, reaching many ($\sim$8) times the effective radii of these objects. The arrows indicate the position of
the effective radii of our galaxies.}

\label{fig2}
\end{figure}

\begin{figure}
%\epsscale{2.0}
\includegraphics[width=\textwidth]{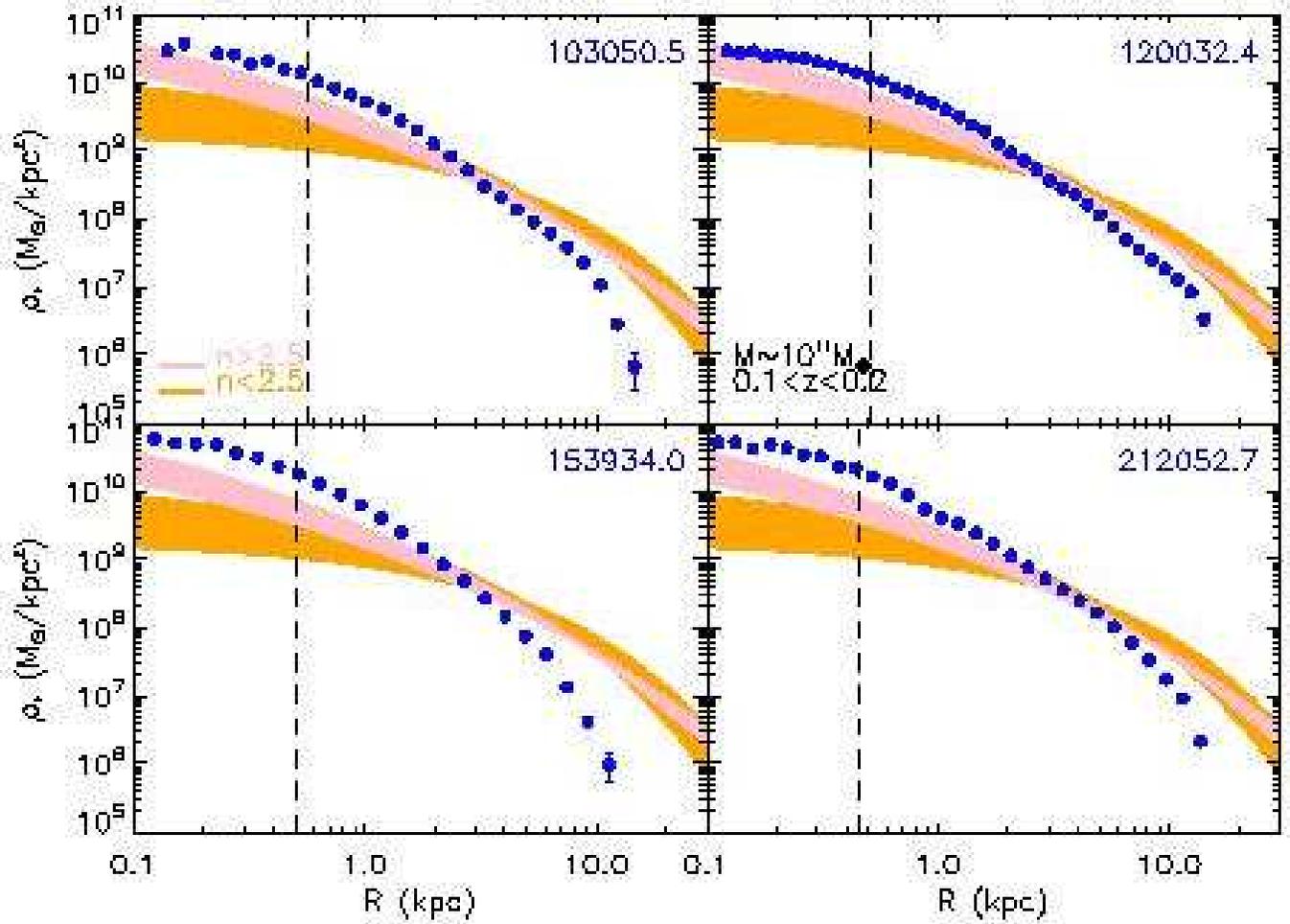}

%\plotone{allmassprofile.ps}

\caption{Stellar surface mass density profiles of our sample of nearby massive compact galaxies (blue
points). The observed profiles of the compact massive galaxies are compared with SDSS DR7 stellar mass density
profiles of M$_\star$$\sim$10$^{11}$M$_{\sun}$ and 0.1$<$z$<$0.2 disk-like galaxies (S\'ersic index
n$<$2.5; orange region)  and  with spheroid-like (S\'ersic index n$>$2.5; pink region)
galaxies. The vertical line shows the equivalent size in kpc of a FWHM PSF of 0.2 arcsec.
The depth and high resolution of our images allow us to explore the profiles of our sample galaxies from
0.1 to 20 kpc.}

\label{fig3}
\end{figure}

\begin{figure}
%\epsscale{2.0}
\includegraphics[width=\textwidth]{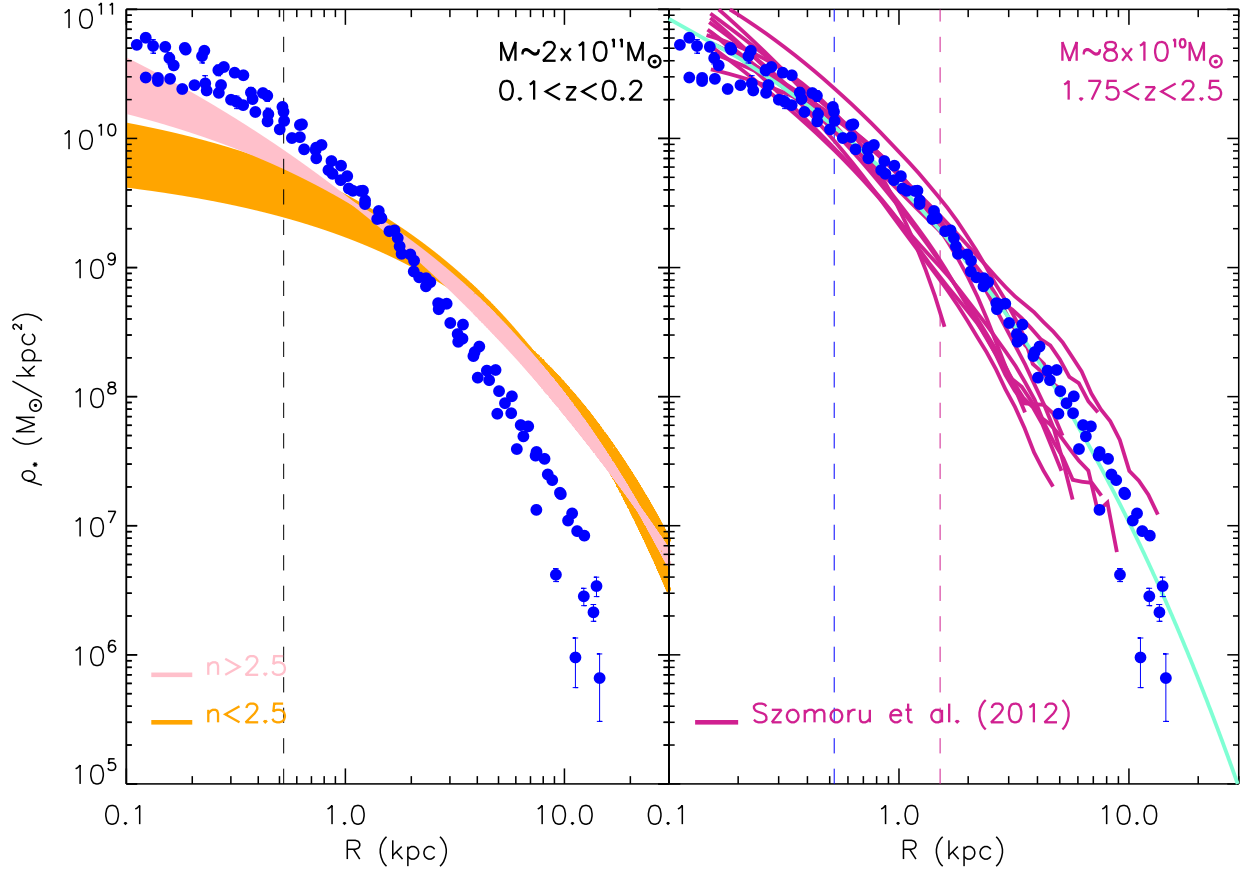}

%\plotone{allmassprofile.ps}

\caption{Stellar surface mass density profiles of our sample of nearby massive compact galaxies (blue
points). {\it Left Panel} The observed profiles of the compact massive galaxies are compared with the
stellar mass density profiles of SDSS DR7 M$_\star$$\sim$2$\times$10$^{11}$M$_{\sun}$ and 0.1$<$z$<$0.2 
disk-like galaxies (S\'ersic index n$<$2.5; orange region) and  with spheroid-like (S\'ersic index
n$>$2.5; pink region) galaxies. The vertical line shows the equivalent size in kpc of a FWHM PSF of 0.2
arcsec at z=0.15. {\it Right Panel} The nearby massive compact galaxies profiles are compared with
profiles of z$\sim$2 massive compact galaxies (violet lines) of the same stellar mass (Szomoru et al. 2012).
The agreement is remarkable. The dashed vertical blue line shows the equivalent size in kpc of a FWHM PSF
of 0.2 arcsec at z=0.15 (our Gemini PSF) and the red vertical line the equivalent size in kpc of a FWHM
PSF of 0.18 arcsec at z=1.9 (HST WFC3 PSF).}

\label{fig4}
\end{figure}

\begin{deluxetable}{cccccccc}
\tabletypesize{\scriptsize}
\tablecolumns{9}
\tablecaption{Galaxy sample and NIRI/Altair observations. \label{data}}
\tablewidth{0pt}
\tablehead{Name   &    R.A.    & Dec       &  Observing date &  Total Exposure time	 &  Airmass & Zero Point  & K mag \\
                  &  (J2000)   & (J2000)   &   (UT)	     &    (sec) 	 &	    &  (AB mag)   & (AB mag)\\ 
          (1)     &    (2)     &    (3)    &	(4)	     &     (5)  	 &     (6)  &	 (7)	  &	(8)	      } 
 \startdata
 SDSS J103050.53$+$625859.8  &  10 30 50.53 & $+$62 58 59.8  & 2010-02-28  & 3540 & 1.406 & 24.86 & 16.63 \\
 SDSS J120032.46$+$032554.1  &  12 00 32.46 & $+$03 25 54.1  & 2010-06-05  & 1620 & 1.106 & 24.85 & 16.06 \\
 SDSS J153934.07$+$441752.2  &  15 39 34.07 & $+$44 17 52.2  & 2010-05-04  & 3660 & 1.281\tablenotemark{a}  & 24.91\tablenotemark{b} & 16.04\\
                             & 		    & 	             & 2010-06-04  &		    &	    &	&	  \\			    
 SDSS J212052.74$+$110713.1  &  21 20 52.74 & $+$11 07 13.1  & 2010-05-22  & 3780 & 1.230 & 24.90  & 16.24\\
\enddata
\tablenotetext{a}{Average airmass from two nights}
\tablenotetext{b}{Average Zero Point for two nights: Night 1: 24.90 mag, Night 2: 24.91 mag}
\tablecomments{Table description - column (1): galaxy name; column (2) and (3): Right Ascension (hours, minutes and seconds) and 
Declination (degrees, minutes and seconds);  column (4): Date of observation in UT; column (5): the  exposure time; column (6): effective airmass; column (7): Derived zero point in the AB system; column (8): K-band galaxy
magnitude}
\end{deluxetable}

\begin{table}
 \centering
 \begin{minipage}{140mm}

  \caption{Nearby massive compact galaxies properties}

  \begin{tabular}{cccccccc}
  \hline
 Name &    M$_\star$           & R$_e$    & S\'ersic index & b/a & R$_e$ & Redshift & $\sigma$   \\
      &    (10$^{10}$M$_\sun$) & (arcsec) & &    & (kpc) &        &  (km/s)    \\
 \hline
 SDSS J103050.53+625859.8 &     8.61 & 0.45$\pm$0.07 & 2.20$\pm$0.24 & 0.36 & 1.23$\pm$0.18 &   0.167 & 196$\pm$16 \\
 SDSS J120032.46+032554.1 &     8.01 & 0.53$\pm$0.08 & 3.55$\pm$0.39 & 0.32 & 1.33$\pm$0.20 &   0.143 & 266$\pm$23 \\
 SDSS J153934.07+441752.2 &     8.79 & 0.38$\pm$0.06 & 2.18$\pm$0.24 & 0.47 & 0.95$\pm$0.14 &   0.143 & 285$\pm$40 \\
 SDSS J212052.74+110713.1 &    10.10 & 0.59$\pm$0.09 & 2.66$\pm$0.29 & 0.28 & 1.35$\pm$0.20 &   0.128 & 223$\pm$12 \\

\hline
\label{data}
\end{tabular}
\end{minipage}
\end{table}
\end{document}